\documentclass[trackchanges, twocolumn]{aastex701}
\usepackage{graphicx}	% Including figure files
\usepackage{amsmath}	% Advanced maths commands
\usepackage{indentfirst}
\usepackage{appendix}
\usepackage{multirow}
\usepackage{booktabs}
\usepackage{threeparttable}
\usepackage{ulem}
\usepackage{soul}
\usepackage{soulpos}

\newcommand{\bi}{\begin{itemize}}
\newcommand{\ei}{\end{itemize}}
\newcommand{\ben}{\begin{enumerate}}
\newcommand{\een}{\end{enumerate}}
\newcommand{\be}{\begin{equation}}
\newcommand{\ee}{\end{equation}}
\begin{document}

\title[Constraining gravity with $DR$]{Constraining gravity with the decay rate of cosmological gravitational potential}

\author{Xinyi Zhao}
\affiliation{Department of Physics, Zhiyuan College, Shanghai Jiao Tong University, Shanghai, 200240, People's Republic of China}
\email[show]{Moon-hits@sjtu.edu.cn}  

\author[0000-0003-2632-9915]{Pengjie Zhang} 
\affiliation{Department of Astronomy, School of Physics and Astronomy, Shanghai Jiao Tong University, Shanghai, 200240, People's Republic of China}
\affiliation{Division of Astronomy and Astrophysics, Tsung-Dao Lee Institute, Shanghai Jiao Tong University, Shanghai, 200240, People's Republic of China}
\affiliation{State Key Laboratory of Dark Matter Physics, Shanghai 200240, People's Republic of China}  
\affiliation{Key Laboratory for Particle Astrophysics and Cosmology (MOE)/Shanghai Key Laboratory for Particle Physics and Cosmology, People's Republic of China}
\email[show]{zhangpj@sjtu.edu.cn}

\author[0000-0003-0296-0841]{Fuyu Dong}
\affiliation{South-Western Institute for Astronomy Research, Yunnan University, Kunming 650500, People's Republic of China}
\email[show]{dfy@ynu.edu.cn}

\begin{abstract}

A key task in cosmology is to test the validity of general relativity (GR)  at cosmological scales and, therefore, to distinguish between dark energy and modified gravity (MG) as the driver of the late-time cosmic acceleration. The decay rate ($DR$) of cosmological gravitational potential, being sensitive to gravity and being immune to various astrophysical uncertainties, enables GR tests independent to other structure growth probes. Recently we have measured $DR$ at $0.2\leq z\leq 1.4$, combining the DR9 galaxy catalog from the DESI imaging surveys and Planck cosmic microwave background maps \citep{Dong25}. Here we use this measurement to test gravity, and restrict the analysis to one-parameter extensions to the standard $\Lambda$CDM cosmology. We consider four one-parameter MG parameterizations. One is $f(a)=\Omega_m^\gamma(a)$. The other three adopt the gravitational slip parameter $\eta=1$ and consider variations in the effective gravitational constant $G_{\rm eff}/G$ with the parameterization $\Sigma(a)=\Sigma_\Lambda \Omega_\Lambda(a)/\Omega_\Lambda$,  $\Sigma(a)=\Sigma_1 a$ or $\Sigma(a)=\Sigma_2 a^2$. We find  $\gamma=0.47^{+0.22}_{-0.15}$, consistent with the GR prediction $\gamma\simeq 0.55$. We also find $\Sigma_\Lambda=0.018^{+0.052}_{-0.053}$, $\Sigma_1=0.020^{+0.065}_{-0.062}$, and $\Sigma_2=0.027^{+0.067}_{-0.069}$, fully consistent with the GR case of $\Sigma=0$, regardless of parameterizations of $\Sigma(a)$. The constraining power is already competitive, while a factor of 2 further improvement is expected for the upcoming full-sky galaxy surveys. 
\end{abstract}

\keywords{\uat{Cosmic microwave background radiation}{322}; \uat{Cosmology}{343}; \uat{Large-scale structure of the universe}{902}}


\section{Introduction} 
A major challenge of modern cosmology is to understand the physics beneath the late-time cosmic acceleration. The main candidate beyond a non-zero cosmological constant $\Lambda$ is a mysterious dark energy (DE) field. A competing possibility is that general relativity (GR) is invalid at cosmological scales and modified gravity (MG) drives the cosmic acceleration \citep{Clifton_2012, Joyce_2015, Ishak_2018, Ferreira_2019}. Some models of the two possibilities can be distinguished by the expansion rate of the Universe, but the rest would require the measurement of structure growth rate to break the degeneracy. 

There are a variety of probes of structure growth rate such as redshift-space distortions (RSD) and weak gravitational lensing \citep{PhysRevD.78.063503, Weinberg_2013}. Among these probes, the integrated Sachs-Wolfe (ISW) \citep{sachs1967} effect is unique in that it directly probes the time variation of the cosmological gravitational potential ($\dot{\psi}+\dot{\phi}$). Here  we adopt the Newtonian gauge and a flat geometry,
\begin{equation}
    d\tau^2=(1+2\psi)dt^2-(1-2\phi)\delta_{ij}dx^idx^j\ .
\end{equation}  
In the linear regime, $\dot{\psi}=\dot{\phi}=0$ for an $\Omega_m=1$ flat universe in which GR is valid. Given the strong observational evidences that our Universe is flat, a non-vanishing ISW effect at large scale implies either the existence of DE or MG. Nonetheless, being a signal overwhelmed by the primary cosmic microwave background (CMB), the measurement of the ISW effect is difficult, and can only be measured through CMB-galaxy (or other tracers) cross-correlations \citep{Crittenden_1996, Nishizawa_2014, 2016A&A...594A..21P}. 

Despite the success of this method, the measured quantity is proportional to the galaxy bias, which weakens its constraining power of cosmology. \cite{ZhangPJ_2006a} proposed the combination of ISW-galaxy and CMB lensing-galaxy cross-correlations into a measurement of the decay rate ($DR$) of the gravitational potential. \cite{Dong22} made the first $DR$ measurement  at $0.2\le z<0.8$ combining DESI imaging surveys and Planck. \cite{Dong25} further extended the measurement to $z\leq 1.2$. 

To constrain cosmology, $DR$ has two advantages. Besides the interested DE and MG parameters, it only depends on the cosmological matter density $\Omega_m$. It does not depend on parameters such as $H_0$, the sound horizon at the drag epoch $r_{\rm d}$, $\sigma_8$ and $n_s$. The other advantage is that, its response to DE  parameters is much stronger than quantities such as $H(z)$ and $D_M(z)$ \citep{Dong22}. Therefore despite the relatively low signal-to-noise ratio (S/N) of $\sim 3$, $DR$ improved over the SDSS dark energy constraint by $30\%$ and the Pantheon dark energy constraint by $40\%$ \citep{Dong22}. Being a structure growth probe,  the constraining power of $DR$ on MG is expected to be more impressive. Therefore in this work we explore its constraints on MG.  The paper is organized as follows. \S \ref{sec.dr} introduces $DR$ and the data sets. \S \ref{sec.modify} presents the MG models adopted in our analysis and the parameter constraints from $DR$. We discuss and conclude in \S \ref{sec.discuss}.

\begin{table}
    \centering
    \begin{tabular}{c|c|c}
    \hline
      Galaxy redshift &$N_g$& $DR$\\
      \hline
      $0.2<z^P<0.4(z_m\simeq0.31)$ & $0.65\times10^6$&$0.047^{+0.066}_{-0.057}$\\
      $0.4<z^P<0.6(z_m\simeq0.51)$ & $1.32\times10^6$&$0.114^{+0.072}_{-0.063}$\\
      $0.6<z^P<0.8(z_m\simeq0.70)$ & $2.47\times10^6$&$0.068^{+0.087}_{-0.060}$\\
      $0.8<z^P<1.0(z_m\simeq0.91)$ & $4.85\times10^6$&$0.105^{+0.087}_{-0.084}$\\
      $1.0<z^P<1.2(z_m\simeq1.09)$ & $3.87\times10^6$&$0.165^{+0.099}_{-0.103}$\\
      $1.2<z^P<1.4(z_m\simeq1.28)$ & $1.33\times10^6$&$0.086^{+0.081}_{-0.081}$\\
    \hline
    \end{tabular}
    \caption{The measured $DR$ \citep{Dong25}. The galaxy samples are selected from DESI imaging survey DR9. }
    \label{table_DR}
\end{table}

\section{The Decay Rate ($DR$) measurement}\label{sec.dr}
The concept of $DR$ is straightforward. The ISW effect $\propto \dot{\psi}+\dot{\phi}$, while weak lensing $\propto \nabla^2(\psi+\phi)$. Therefore combining the ISW-galaxy cross-correlation and the weak lensing-galaxy cross-correlation, we  are able to measure the following combination \citep{ZhangPJ_2006a}, 
\begin{equation} \label{eq:dr}
    DR(z)=\left(-\frac{d\ln{D_{\psi+\phi}}}{d\ln{a}}\right)\left(\frac{aH(z)/c}{W_L(z)}\right)\ .
\end{equation}
Here $W_L(z)=[1-\chi(z)/\chi(z_s)]/\chi(z)$ is the lensing kernel, and $\chi(z)$ is the comoving radial distance to redshift $z$. $z_s$ is the source redshift. In the case of CMB lensing, $z_s\simeq 1100$. $D_{\psi+\phi}$ is the linear growth factor of the (Weyl) potential. From the definition, $DR$  is independent of galaxy bias, the matter clustering amplitude and slope, and the Hubble constant $H_0$. Hence $DR$ is significantly more sensitive to the nature of dark energy and gravity than other probes, in case of identical measurement precision \citep{Dong22}.  

$DR$ measured using DESI imaging surveys DR9 and Planck is listed in Table \ref{table_DR} \citep{Dong25}.  In order to construct a volume-limited sample, galaxies are further selected from the DECaLS+DES region with z-band absolute magnitude brighter than -22.  To mitigate imaging systematics in $\delta_g$, each galaxy is assigned with a weight derived from a machine learning method \citep{Xu2023}. Furthermore, the magnification bias was mitigated using various methods. The total significance of the $DR$ measurements is $3.1\sigma$. Due to the built-in sensitivity of $DR$ to gravity, we will show that even a $3\sigma$ measurement delivers competitive constraints on MG.

\begin{figure}
    \centering
    \includegraphics[width=.46\textwidth]{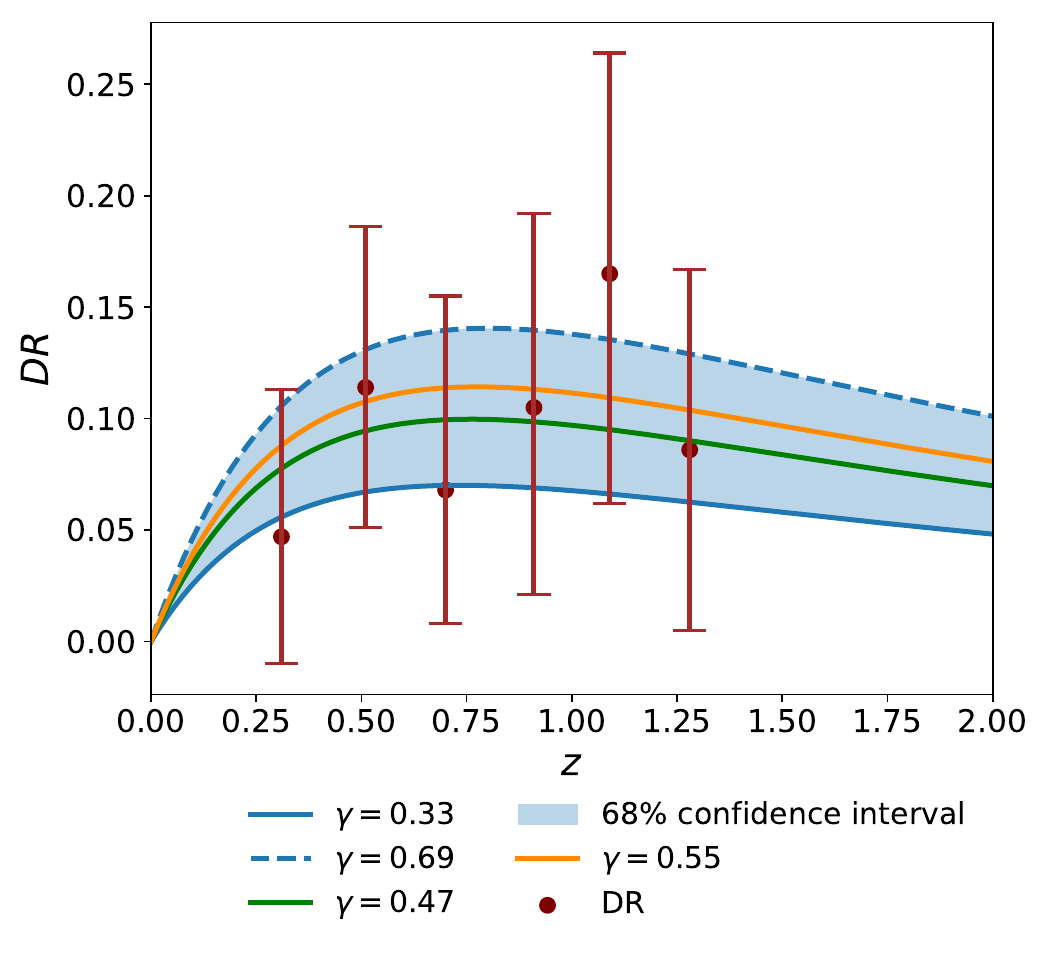}
    \caption{Constraint on the $\gamma$ parameter using $DR$. The shaded areas indicate the $68\%$ confidence region.   $\gamma=0.47_{-0.15}^{+0.22}$, consistent with the GR value of $\gamma\simeq 0.55$. \label{fig.gammas_conf}}
\end{figure}
\section{Constraints on Modified Gravity}\label{sec.modify}
To test MG, we restrict our analysis to one-parameter extension to the flat $\Lambda$CDM cosmology. Within this setup, $DR$ depends only on $\Omega_m$, other than the gravity parameters. $\Omega_m$ has been measured to high precision using probes of the expansion rate, such as the peak position of CMB, baryon acoustic oscillation (BAO) and supernovae (SNe).  We adopt $\Omega_m=0.3169\pm0.0086$ \citep{Desi_dr2_result2}, combining Planck, the Atacama Cosmology Telescope (ACT), and the South Pole Telescope (SPT). Later we will show that the $0.3\%$ statistical uncertainties in $\Omega_m$  impacts on constraints of MG parameters at the level of $\la 0.1\sigma$. Therefore we will neglect the above statistical uncertainties in $\Omega_m$ and fix $\Omega_m=0.3169$ throughout the paper, except in \S \ref{sec.discuss} where we will evaluate the impact of the $\Omega_m$ tension between BAO, CMB and SNe Ia. We consider 4 different one-parameter parameterizations of MG.

\begin{figure}
    \centering
\includegraphics[width=0.46\textwidth]{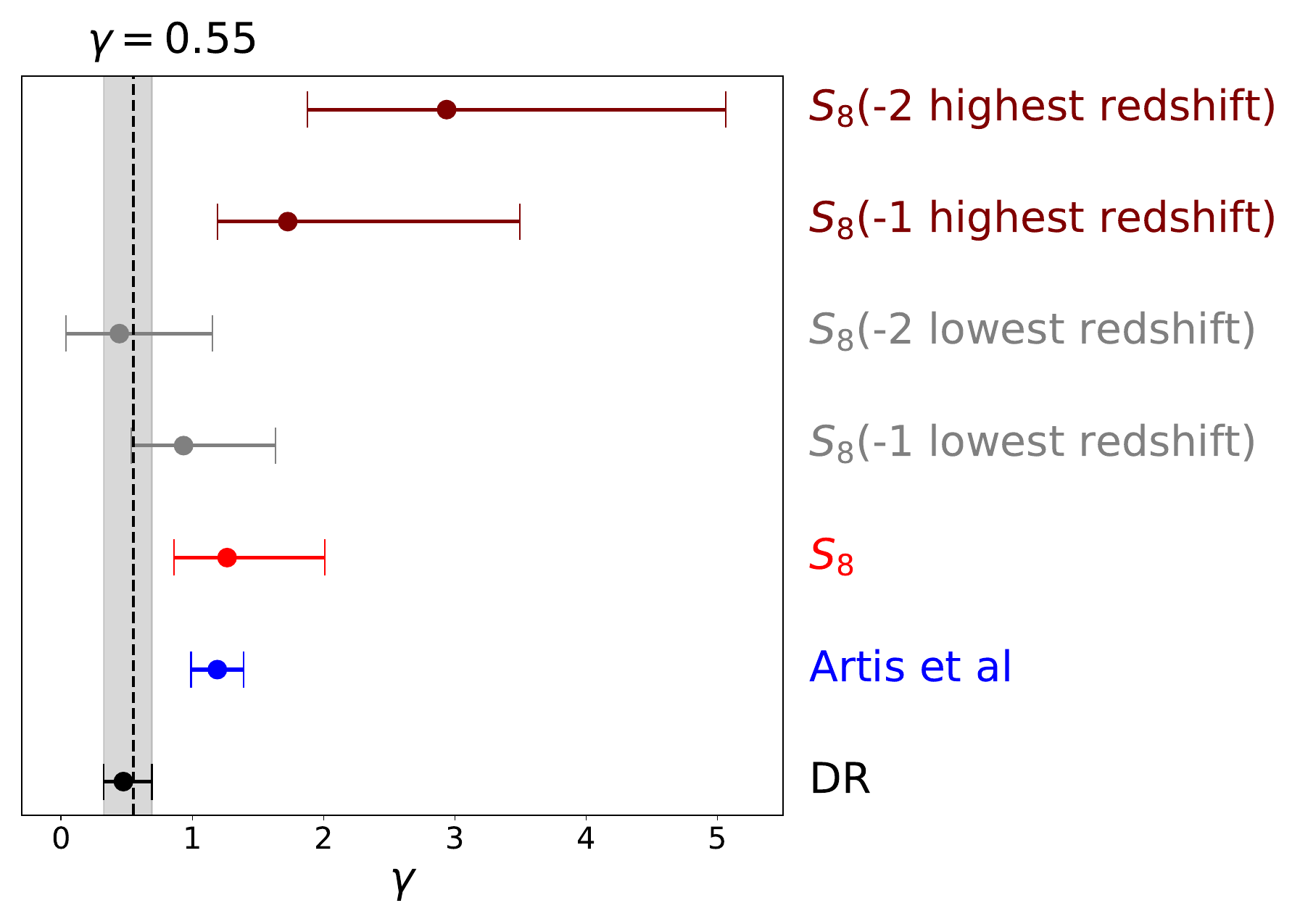}
\caption{$\gamma$ constraints from various data sets. In the bottom is $\gamma$ constrained by $DR$. The constraint labeled with "Artis et al" is from \cite{Artis24} using  cluster number counts. The one with label "$S_8$" is calculated by all the $S_8(z)$ measured by Fig.19 of \cite{Qu24} (includes $S_8(z)$ measurements of \cite{Qu24,Sailer24,Farren_2024} ). The other constraints are calculated by disregarding one or two data points.  \label{fig.S8comparison}}
\end{figure}
\subsection{$DR$ constraint on the $f=\Omega^\gamma_m(a)$ model}
The first MG parameterization that we consider is the $\gamma$-parameterization for the structure growth rate $f\equiv d\ln D/d\ln a$. Here $D$ is the linear density growth rate. $f=\Omega^\gamma_m(z)$ \citep{1980lssu.book.....P, 1985PhLB..158..211F, 1990ApJS...74..831L, Wang_1998, Linder_2005}. For $\Lambda$CDM around the best-fit cosmological parameters, $\gamma\simeq 0.55$. Therefore $\gamma\neq  0.55$ indicates the departure of gravity from GR. For example, $\gamma=11/16\simeq 0.68$ for the Dvali–Gabadadze–Porrati (DGP) model in the self-accelerating branch \citep{Lue_2004b, Linder_2005}. Note that since both ISW and weak lensing measure $\psi+\phi$ instead of the matter overdensity $\delta$, we assume that the $(\psi+\phi)$-$\delta$ relation remains the same as in GR. This simplification is applicable to some modified gravity models such as DGP \citep{Lue_2004b} and $f(R)$ \citep{ZhangPJ_2006b}. 

Fig.\ref{fig.gammas_conf} shows the results. A larger $\gamma$ causes weaker structure growth, stronger decay of $\psi+\phi$, and hence larger $DR$.  We find that $\gamma=0.47_{-0.15}^{+0.22}$ (also shown in Table \ref{tab.result}), fully consistent with the GR prediction $\gamma\simeq 0.55$. 

Our result of $\gamma=0.47_{-0.15}^{+0.22}$, using independent data and method, provides a different viewpoint on recent controversies in the $\gamma$ determination.
For example, \cite{Artis24} found $\gamma=1.19\pm 0.2$ using X-ray cluster number counts of eRASS1. Given $S_8(z)$ measured by CMB lensing-galaxy cross-correlations, we can also infer $\gamma$.  We combine the $S_8(z)$ measured by \cite{Farren_2024, Qu24,Sailer24} using various combinations of CMB lensing (Planck/ACT) and galaxies (DESI/unWISE). The resulting constraint on $\gamma$ is shown in Fig. \ref{fig.S8comparison}. The data prefer $\gamma\sim 1.2$ or even larger, unless the data point in the lowest redshift bin ($z\sim 0.37$) is excluded. Note that the two constraints on $\gamma$ obtained from X-ray cluster number counts and CMB lensing may appear to be consistent, but this is not the case. Using Planck's measurement $S_8$ ($\sigma_8$) as a reference, the $\gamma\sim 1.2$ value of the eROSITA eRASS1 result is driven by a higher $\sigma_8$ at $z \sim 0.6$. However, in the CMB lensing case, the same $\gamma\sim 1.2$ originates from a lower $S_8$ at $z\sim 0.37$.

\begin{table}
    \centering
    \resizebox{\linewidth}{!}{\begin{tabular}{ccc}
    \toprule
    Parameter&Best-fit &MG Model\\
         \midrule
         $\gamma$ & $0.47_{-0.15}^{+0.22}$ & $f=\Omega_m(a)^\gamma$  \\
         %$\gamma$ & $0.48\pm0.16$ & $0.16$ & $f=\Omega_m(a)^\gamma$  \\
         $\Sigma_\Lambda$ & $0.018^{+0.052}_{-0.053}$  & $\Sigma=\Sigma_\Lambda\frac{\Omega_\Lambda(a)}{\Omega_{\Lambda}}$ \\
         $\Sigma_1$ & $0.020^{+0.065}_{-0.062}$ & $\Sigma=\Sigma_1 a$ \\
         $\Sigma_2$ & $0.027^{+0.067}_{-0.069}$ & $\Sigma=\Sigma_2 a^2$ \\
    \bottomrule
    \end{tabular}}
    \caption{Best-fit values of the MG parameters and the $1\sigma$ errors.}
    \label{tab.result}
\end{table}
 
\subsection{Constraints on $G_{\rm eff}$}
 At linear perturbation level and sub-horizon scale, there are two degrees of freedom to modify GR. Among various equivalent parameterizations (e.g., \cite{Zhang07, Caldwell_2007, PhysRevD.76.023514, Amendola_2008, Zhao_2009, Song_2009,  Bean_and_Tangmatitham_2010}), one is the ratio $\eta\equiv \phi/\psi$ and the other is the effective Newton's constant in $k^2(\psi+\phi)=-8\pi G_{\rm eff} a^2\rho \delta$. The ratio $G_{\rm eff}/G$ is often parameterized as $G_{\rm eff}/G=1+\Sigma(a)$.  The evolution of $\delta$ is determined by $\psi$ and the $\psi$-$\delta$ relation is now
\begin{equation}\label{eq.poisson0}
k^2\psi=-8{\pi}Ga^2\rho\delta \frac{1+\Sigma}{1+\eta}\ .
\end{equation}
The $DR$ data alone is not able to break the degeneracy between $\eta$ and $\Sigma$. Therefore, we will fix $\eta=1$ and only consider $\Sigma\neq 0$. Under this simplification,  Eq. \ref{eq.poisson} becomes
\begin{equation}\label{eq.poisson}
k^2\psi=-4{\pi}Ga^2\rho\delta (1+\Sigma)\ .
\end{equation}
Then $DR$ becomes
\begin{equation}\label{eq:dr_sigma}
    DR = \left(1 - f - \frac{d\ln{(1+\Sigma)}}{d\ln{a}}\right)\frac{aH(z)/c}{W_L (z)}\ .
\end{equation}
We numerically solve the following equation to determine $f$,
\begin{equation}\label{eq:linearp}
    \delta^{\prime\prime}+\delta^\prime (\frac{H^\prime}{H}+\frac{3}{a})-
    \frac{3}{2}\frac{\Omega_m H_0^2}{H^2 a^3}\frac{\delta}{a^2}(1+\Sigma)=0\ .
\end{equation}
Here $'$ means $d/da$.

\begin{figure}
    \centering
\includegraphics[width=.46\textwidth]{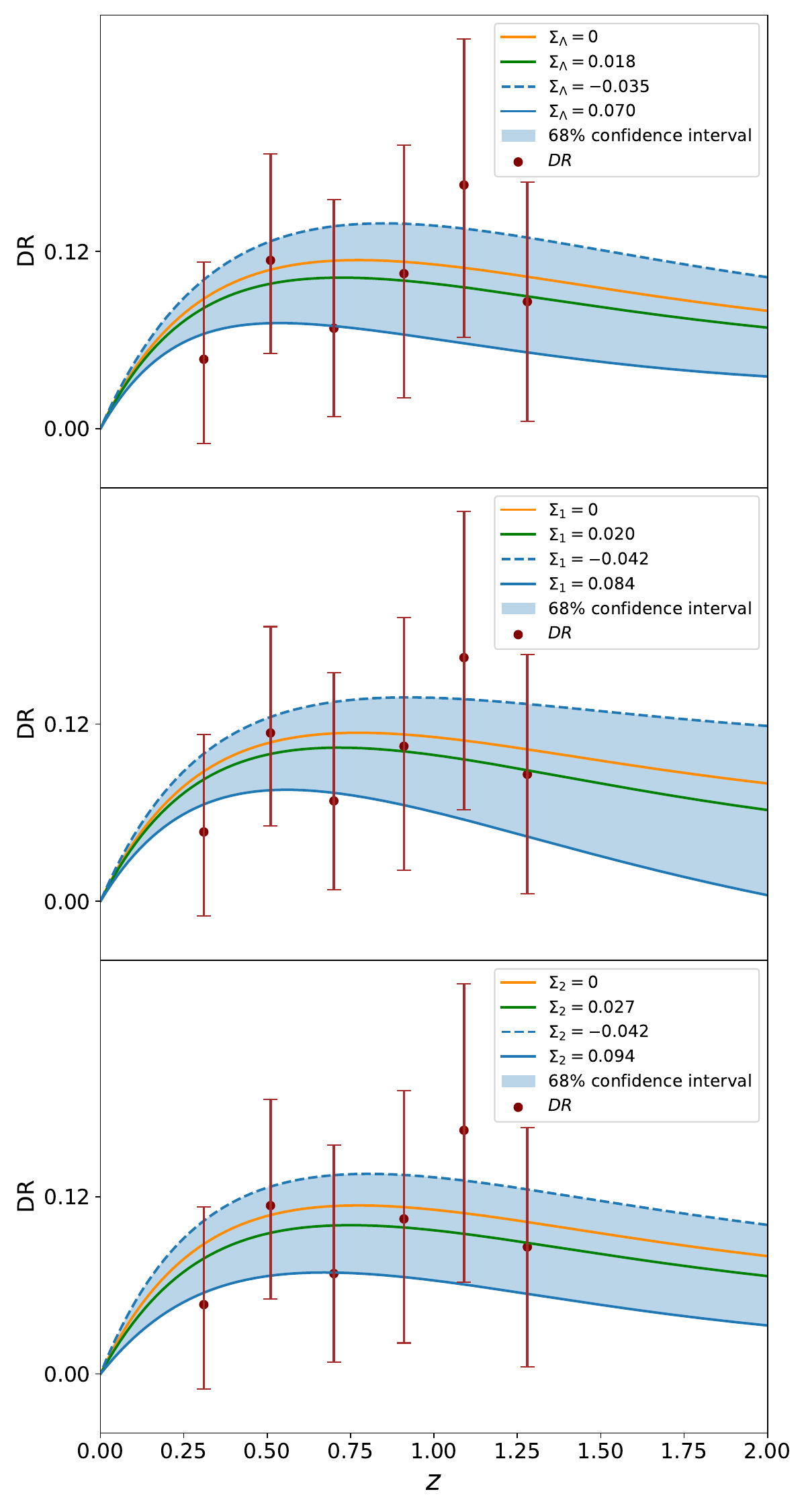}
    \caption{Similar to Fig.\ref{fig.gammas_conf}, but for the $\Sigma_\Lambda$, $\Sigma_1$ and $\Sigma_2$ parameterizations.}
    \label{fig.diff_model}
\end{figure}
We need to specify the redshift dependence of $\Sigma$. Since cosmic acceleration occurred at late times, we naturally expect $\Sigma\rightarrow 0$ when $a\rightarrow 0$. Then we adopt the following three forms of $\Sigma(a)$.
\begin{itemize}
\item  $\Sigma(a)=\Sigma_\Lambda\Omega_\Lambda(a)/\Omega_\Lambda$. This is widely adopted in MG tests (e.g., \citet{2016A&A...594A..14P,Abbott_2019,2020A&A...641A...6P,2021PhRvD.103h3533A,2025JCAP...09..053I}). $\Omega_\Lambda(a)$ is the dark energy density at redshift $z=1/a-1$.
\item $\Sigma(a)=\Sigma_1 a$. The evolution is faster at $a=1$, but weaker at $a\rightarrow 0$, than $\Sigma(a)=\Sigma_\Lambda\Omega_\Lambda(a)/\Omega_\Lambda$. 
\item $\Sigma(a)=\Sigma_2 a^2$. The evolution is also faster at $a=1$, but weaker at $a\rightarrow 0$, than $\Sigma(a)=\Sigma_\Lambda\Omega_\Lambda(a)/\Omega_\Lambda$. 
\end{itemize}
 In the following, we will use the symbol $\Sigma_X$ to represent the parameter $\Sigma_\Lambda|\Sigma_1|\Sigma_2$. A positive $\Sigma_X$ speeds up the structure growth and decreases $DR$, while a negative $\Sigma_X$ results in a larger $DR$.

We show the fitting results  in Fig. \ref{fig.diff_model}, and also in Table \ref{tab.result}. For example, ${\Sigma_\Lambda}=0.018^{+0.052}_{-0.053}$. Both the best-fit values and the associated error bars depend on the parameterization of $\Sigma(a)$. However, despite these differences, in all three cases $\Sigma_X$ is consistent with zero (GR) within $\sim 0.5\sigma$. Such consistency  provides a further support of GR, regardless of the $\Sigma(a)$ parameterization. 

It is worth noting that the constraint using $DR$ ($\sigma_{\Sigma_\Lambda}\sim 0.05$) is already as tight as that combining DESI full shape galaxy clustering, Planck CMB, DESY3 weak lensing and DES supernovae (Fig. 3 \& 4 of \cite{2025JCAP...09..053I}). Note that since we fix $\eta=1$, a fair comparison should be restricted  to their posterior distribution of $\mu_0$ with fixed $\eta_0=0$. This demonstrates the competitive constraining power of the $DR$ measurement. In future work we will combine  other LSS probes  to break the $G_{\rm eff}$-$\eta$ (or $\Sigma$-$\eta$, $\mu$-$\Sigma$) degeneracy.

\section{Discussions and Conclusion}\label{sec.discuss}
\begin{figure}
    \centering
    \includegraphics[width=0.47\textwidth]{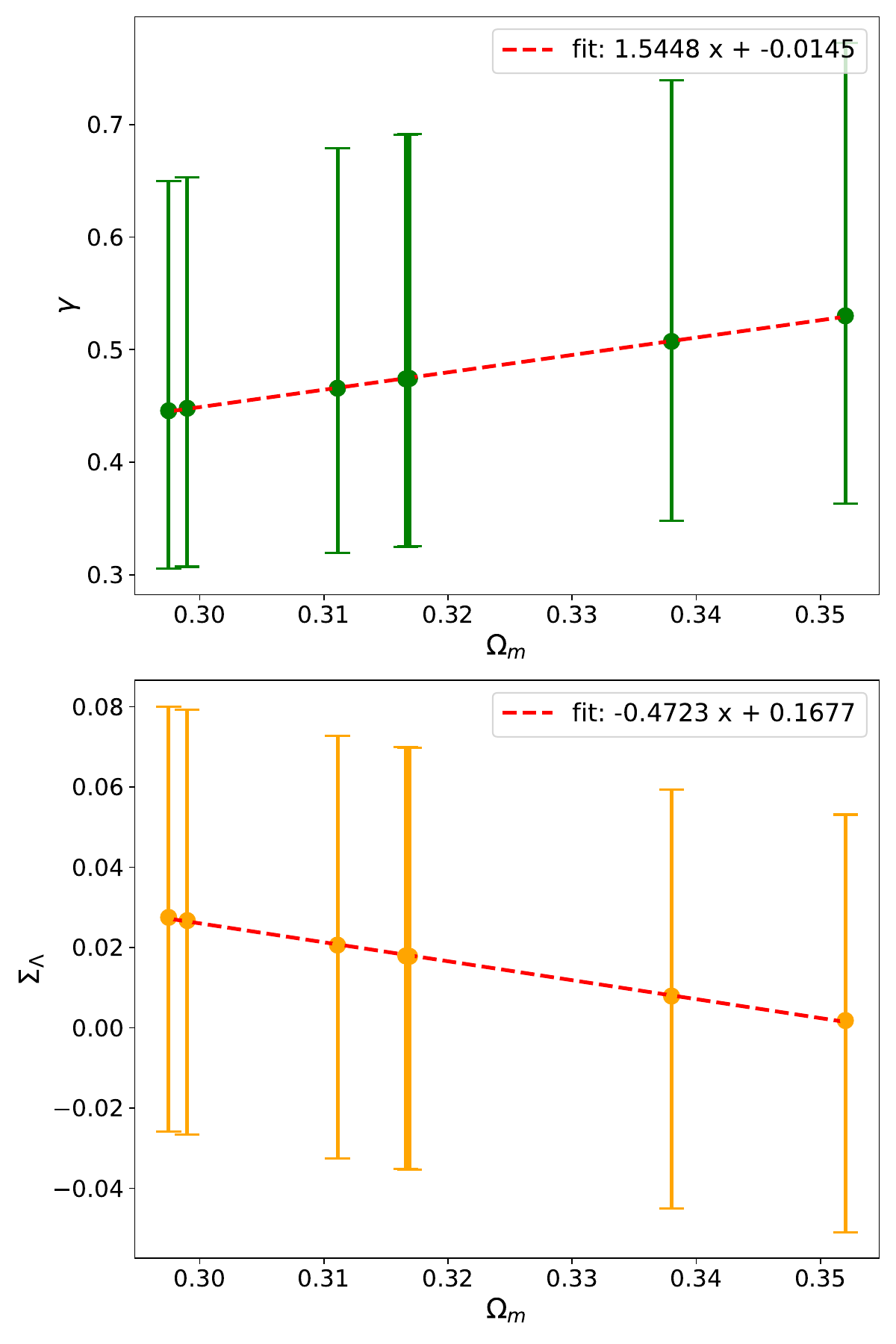}
    \caption{The dependence of MG parameter constraints on $\Omega_m$. Over the  range of best-fit $\Omega_m$ from BAO, CMB and SNe Ia, the induced variation in $\gamma$ and $\Sigma_\Lambda$ is subdominant to the statistical error, and the agreement with GR remains unchanged.  }
    \label{fig_Omega_gm}
\end{figure}

We have showed that the $DR$ measurement supports GR, using 4 parameterizations of MG models. For brevity, we have fixed $\Omega_m=0.3169$, as constrained from CMB. Fig. \ref{fig_Omega_gm} shows the dependence of $\gamma$ and $\Sigma_\Lambda$ constraints on $\Omega_m$. The $0.3\%$ uncertainty in the CMB $\Omega_m$ constraint translates into a $\la 0.1\sigma$ shift in the best-fit value of $\gamma$, and $\Sigma_\Lambda$. Therefore our previous simplification of fixing $\Omega_m$ is justified.  However, differences in the best-fit $\Omega_m$ from BAO, CMB and SNe Ia are much larger, ranging from $\Omega_m=0.2975$ (BAO) \citep{Desi_dr2_result2} to $\Omega_m=0.352$ (DES5Y SNe Ia) \citep{2024ApJ...973L..14D}. The induced variation in the best-fit $\gamma$ ($\Sigma_\Lambda$), along with the associated $1\sigma$ errorbars are shown in Fig. \ref{fig_Omega_gm}. The dependence on $\Omega_m$ is well approximated to be linear. Switching from the CMB best-fit $\Omega_m$ to BAO/SNe best-fit $\Omega_m$, the shift in the best-fit $\gamma$ and $\Sigma_\Lambda$ is smaller than $0.3\sigma$. Therefore for the current data quality, such uncertainties induced by this $\Omega_m$-tension are negligible, and the agreement with GR is unaffected. 

\begin{figure}
    \centering
\includegraphics[width=.44\textwidth]{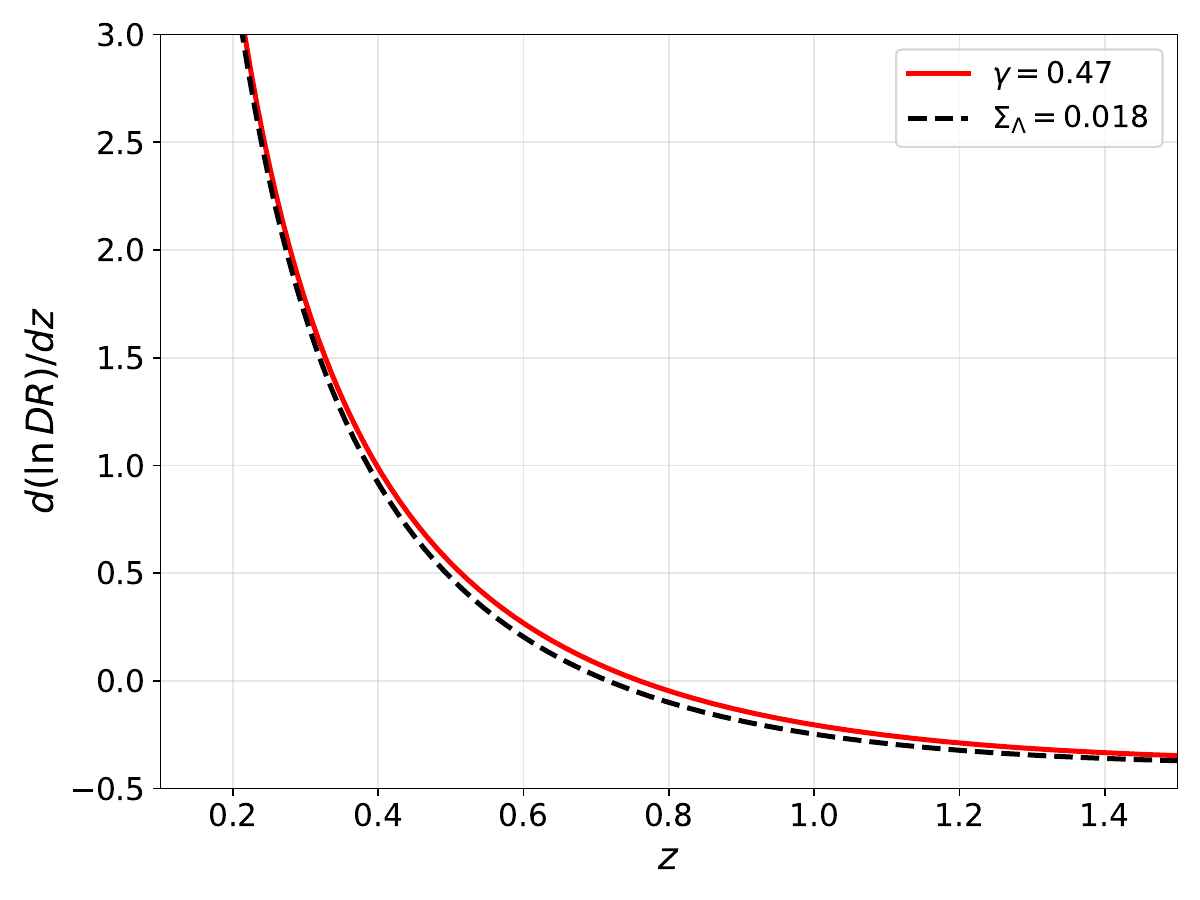}
    \caption{Redshift sensitivity of $DR$ quantified by $d\ln{DR}/dz=(d DR/dz)/DR$. The red line is calculated from the derivative of Eq.\ref{eq:dr} given our best-fit $\gamma=0.47$. The black dashed line is calculated from Eq.\ref{eq:dr_sigma} using the form $\Sigma(a)=\Sigma_\Lambda\Omega_\Lambda(a)/\Omega_\Lambda$, and using the best-fit $\Sigma_\Lambda=0.018$. The relativistic $DR$ bias is less than $3\%$ given the redshift calibration bias in Table \ref{table_DR}, well below the statistical uncertainty of the $DR$ measurements.
    \vspace{2pt}
    }
    \label{fig:dlnDR_dz}
\end{figure}

We have also tested the impact of possible redshift calibration uncertainties in the $DR$ measurements. In our fiducial analysis, each data point is modeled at the mean redshift $z_m$ listed in Table \ref{table_DR}. Here we show $d\ln{DR}/dz$ in Fig.\ref{fig:dlnDR_dz} to demonstrate this calibration has little effect on $DR$. We would expect the systematic bias in the mean redshift $z_m$ is  $|\delta z| \sim 0.02$. The induced fractional bias in $DR$ is $3\%$ at $z=0.3$ and below $\%$ at $z>0.5$. This bias is significantly below the $DR$ measurement uncertainties. Therefore, redshift calibration is not a problem for the current cosmological application of $DR$. As a further robustness test, we repeated the parameter constraints after dropping either the lowest or the highest redshift bin, as these bins may have larger bias in the mean redshift $z_m$. Dropping the lowest redshift bin gives $\gamma=0.51^{+0.24}_{-0.16}$ and $\Sigma_\Lambda=0.009^{+0.054}_{-0.056}$, while dropping the highest redshift bin gives $\gamma=0.48^{+0.25}_{-0.17}$ and $\Sigma_\Lambda=0.016^{+0.059}_{-0.061}$. These results are fully consistent with the fiducial analysis. Therefore we conclude that our constraints on gravity parameters are robust. 

This work demonstrates the impressive power of $DR$ to constrain MG. Therefore a major task in this direction is to improve the $DR$ measurement precision and accuracy. Upcoming full sky galaxy surveys with wide redshift coverage will improve the ISW measurement and hence the $DR$ measurement by a factor of $2$. Meanwhile, surveys such as DESI and the planed stage IV spectroscopic redshift surveys, with accurate redshifts and well-controlled imaging systematics, will further reduce potential systematics in the $DR$ measurement. Together with other probes such as redshift space distortion, we expect significant improvement in constraints on both $G_{\rm eff}$ and $\eta$ (or equivalently $\mu$ and $\Sigma$). Given the strength in the $DR$ constraining power and the simplicity in the $DR$ data/modeling, it is beneficial to include $DR$ in observational tests of gravity.

\section{acknowledgements}
P.Z. and X.Z. are supported by  the National Key R\&D Program of China (2023YFA1607800, 2023YFA1607801).
F.D. is supported by  the National Natural Science Foundation of China (grant No.12303003).

\bibliography{citations}{}
\bibliographystyle{aasjournalv7}



\end{document}